\documentclass[printer]{aa}

\usepackage[varg]{txfonts}
\usepackage{color}
\usepackage{graphicx,subfigure}
\usepackage{amsmath}
\usepackage{epsfig,amsmath,amssymb}

\def\be{\begin{equation}}
\def\ee{\end{equation}}  
\def\ba{\begin{eqnarray}}
\def\ea{\end{eqnarray}}  
\def\ro{R_{\rm out}}

\newcommand{\calpha}{C_\alpha}
\newcommand{\cbeta}{C_\beta}

\def\fig#1{fig.~(\ref{#1})}

\def\s0#1#2{\mbox{\small{$ \frac{#1}{#2} $}}}
\def\0#1#2{\frac{#1}{#2}}

\begin{document}

%
\titlerunning{Influence of coronal currents on $\alpha^2$ stellar dynamo}
\authorrunning{Bonanno, Del Sordo}
\title{An analytic  mean-field $\alpha^2$-dynamo with a force-free corona}
\author{Alfio Bonanno\inst{1} and Fabio Del Sordo\inst{2,3}}
\date{Received -- / Accepted --}

\institute{
INAF Osservatorio Astrofisico di  Catania\\
\email{alfio.bonanno@oact.inaf.it}
\and
Department of Geology \& Geophysics, Yale University, New Haven, USA
\and
Nordita, Royal Institute of Technology and Stockholm University, SE-10691 Stockholm, Sweden \\
\email{fabio@nordita.org} \\
}

\abstract
{
Stellar dynamos are affected by boundary conditions imposed by stellar coronae. 
Under some approximations it is possible to find analytical solutions. 
Interior dynamo models often consider a current-free coronae without taking into account
the constraints imposed by the presence of currents in the corona. 
}
{
We aim to analytically evaluate the effect of coronal currents and of an outer boundary condition  on the efficiency 
of an $\alpha^2$ dynamo. 
We intend to estimate the change in geometry and dinamo excitation numbers with respect to the current-free case.
}
{
We analytically solve the turbulent dynamo induction equation  for a homogeneous, non-mirror symmetric turbulence, 
in a spherical domain surrounded by a linear force-free corona with mean magnetic field $\mathbf{B}$ satisfying $\nabla\times \mathbf{B}= \beta \mathbf{B}$.
}
{
The dynamo number is a decreasing function of $\beta$. Moreover, 
if the current is parallel to the field ($\beta>0$) the dynamo number is smaller than in the force-free case. On the contrary ($\beta<0$) 
the dynamo number is greater than in the force-free case.  
}
{
The presence of currents in the corona needs to be taken into account because it affects the condition for excitation of a dynamo.
}


\maketitle
\section{Introduction}
A consistent description of the magnetic coupling between a magnetized coronal field and a dynamo generated interior model 
is still an open issue. 
Mean field models usually employ a current-free description of the coronal field \citep{krause80, Moffatt80}, an assumption which is not compatible with an active corona. 

However, direct numerical simulations are not able to resolve the surface layers where the pressure scale height is very small and the Lorentz force  cannot be neglected.
Nonetheless, recent models have been used to estimate the effect of a nearly force-free corona on 
controlling the emergence of flux from lower layers \citep{WBM2011}, on the injection of magnetic twist in the heliosphere \citep{WBM2012} as well as on a global dynamo \citep[e.g.][]{Warnecke2016}, concluding that the presence of a corona cannot be neglected.

Common models of coronal field based on a potential field  \citep{nash88}
\be
\nabla^2 \mathbf{b}=0   
%
\ee
or on a  force-free field 
\be\label{ff0}
\nabla\times\mathbf{b}=\beta \, \mathbf{b} 
%
\ee
where $\mathbf{b}$ is the total magnetic field, $\beta=\beta(\mathbf{x})$ is a scalar function, 
are often used to describe chromospheric observations \citep{weige2012}. On the other hand the coupling with the interior is always neglected.
In the framework of mean-field dynamo theory a new proposal has been presented in \cite{bo16} where a consistent
coupling with an $\alpha^2\Omega$ dynamo model for the interior has been achieved under the assumption that the coronal
field is harmonic. On the other hand, there are no known analytical solutions in spherical symmetry of a dynamo generated interior field 
coupled with a force-free exterior. The aim of this paper is to present such a solution for a linear force-free field. 
Although our solution has been obtained for  a very idealized case (non-helical homogeneous turbulence), 
some of its features are in agreement with the findings of \cite{bo16}. 

Moreover, as it is well known, in order to obtain the photospheric field from Zeeman-Doppler Imaging (ZDI) is 
it is necessary to solve strongly non-linear inverse problem which is very often beset by the presence of several local minima.
The choice of a proper forward model is essential to restrict the space of possible solutions to the physical ones
\citep{ca09,stift17}. We therefore hope that our analytical solution can be useful also in the ZDI reconstruction approach. 

The structure of this article is the following: 
Sect. 1 contains the physical motivation of this work,
Sect. 2 contains the basic equations of the model and their physical implications,
and Sect. 3 is devoted to the conclusions. 

\section{The model}
\subsection{Basic equations}
Let us assume that the field in the stellar interior admits a description in terms of the standard mean-field dynamo equation.

In the case of homogeneous, isotropic, and non-mirrorsymmetric turbulence,
we can write the evolution equation for the mean magnetic field $\mathbf{B}$ 
\be\label{induction}
\frac{\partial\mathbf{B}}{\partial t} = \nabla\times(\mathbf{U}\times\mathbf{B}+\alpha \mathbf{B})-
\nabla\times(\eta\nabla\times\mathbf{B})
\ee
where $\alpha$ describes the turbulent flow, $\mathbf{U}$ is the mean flow and $\eta$ characterizes the turbulent (eddy) diffusivity.  $\alpha$ and $\eta$ are simple coefficients in our case, but they would be tensors if isotropy were not assumed. A mean field in a turbulent medium is defined as the expectation value of the total field in an ensemble of identical systems. In our case, the mean magnetic field can be seen as a time average of the total magnetic field over a time scale that is short if compared to the long term evolution of the field.

If $\mathbf{U}=0,$ we have $\alpha=const$, $\eta=const$, so
an exact stationary  solution can be obtained by employing the standard  decomposition in toroidal and poloidal component
\be\label{deco}
\mathbf{B}=-\mathbf{r} \times \nabla \Psi - \nabla\times (\mathbf{r}\times \nabla \Phi)\equiv
\mathbf{B}_{\rm T}+\mathbf{B}_{\rm P}
\ee

where $\Psi=\Psi(r,\vartheta,\varphi)$ and 
$\Phi=\Phi(r,\vartheta, \varphi)$ are scalar functions,  $\mathbf{B}_{\rm T}$ is the toroidal component and $\mathbf{B}_{\rm P}$ is the poloidal one (see \cite{krause80} for details). 
By inserting \eqref{deco} in \eqref{induction} one easily finds

\begin{subequations}
\label{a2}
\ba
&&\frac{\alpha}{\eta}\Psi+\nabla^2 \Phi=0, \\ \label{a2t}
&&\nabla^2 \left (\frac{\alpha}{\eta}\Phi-\Psi\right )=0
\ea
\end{subequations}
which implies 
\be
\label{torin}
\nabla^2\Psi+ \left ( \frac{\alpha}{\eta}\right )^2 \Psi=0
\ee
It is convenient to employ the following decomposition
\begin{subequations}
\label{d1}
\ba
&&\Phi=R \sum_{n=1}^\infty\sum_{m=-n}^{m=n} \phi_{mn}(x)Y_n^m(\theta,\varphi) \\ \label{tin}
&&\Psi=\sum_{n=1}^\infty\sum_{m=-n}^{m=n}\psi_{mn}(x)Y_n^m(\theta,\varphi)
\ea
\end{subequations}
where $x$ is the normalized stellar radius $x=r/R$,
$\phi_{nm}(x)$ and $\psi_{nm}(x)$ are the eigenfunctions of the radial elliptic problem defined by \eqref{a2},
$Y_n^m(\vartheta, \varphi)$ are the spherical harmonics functions. 

By inserting \eqref{tin} in \eqref{torin} we obtain that the only regular interior solution is 
\be
\label{soltin}
\psi_{nm}(x)= A_{nm} \frac{J_{n+1/2}(\calpha x)}{\sqrt{x}}
\ee
where $J_n(x)$ is the  Bessel function of the first kind, and $A_{nm}$ is a complex coefficient which must be determined assuming an exterior field configuration. 
For simplicity's sake we will use $A_{nm}=1$ whenever we will need to normalize quantities such as mean magnetic field $\mathbf{B}$ and magnetic helicity $\mathbf{A}\cdot\mathbf{B}$.
By substituting \eqref{soltin} in \eqref{a2t} one finds \citep{krause80} 
\be
\label{solsin}
\phi_{nm}(x)= A_{nm} \frac{J_{n+1/2}(\calpha x)}{\calpha \sqrt{x}}+ B_{nm} x^n
\ee  
where $B_{nm}$ is a constant, and $\calpha=\alpha / \eta$ is the eigenvalue parameter of \eqref{a2}.

\subsection{Adding a force-free corona}
We now consider that the dynamo domain is surrounded by a corona, therefore
in the exterior it is assumed that the field satisfies linear force-free condition, namely 
\eqref{ff0} with $\beta=const$. 
This is the state of minimum dissipation for a given amount of magnetic energy \citep{ChandraWoltjer1958}.
As explained in \cite{chandra57}, the general solution of \eqref{ff0} can be obtained 
by inserting \eqref{deco} in \eqref{ff0}, thus obtaining the Helmholtz equation for the $\Phi$ function
\be\label{ffh}
\nabla^2\Phi+\beta^2 \Phi=0
\ee
and $\Psi=\beta\Phi$ in this case.  The general solution of \eqref{ffh} reads
\be
\label{solff}
\Phi=R \sum_{n=1}^\infty\sum_{m=-n}^{m=n}[C_{nm}j_n{(\cbeta x)}+D_{nm}y_n(\cbeta x)]Y_n^m(\theta,\varphi) 
\ee
where $\cbeta=\beta R$, $C_{nm}$ and $D_{nm}$ are coefficients depending on n and m, 
$j_n(x)$ and $y_n(x)$ are the spherical Bessel functions:
\be
j_n(x) = \sqrt{\frac{\pi}{2 x}} J_{n+1/2}(x), 
\;\;\; y_n(x) = \sqrt{\frac{\pi}{2x}}Y_{n+1/2}(x)
\ee
being $J_n(x)$, $Y_n(x)$ the  Bessel function of the first and second kind, respectively. 
Notice that, as $r\geq R$, differently from what discussed in 
 \cite{naka73} and \cite{priest82}, the general solution of the linear force-free equations must include both first kind and second kind Bessel's functions.

We impose the continuity of all the field components across the boundary so that 
\be
[[\mathbf{B}]]=0 ,
\ee
where the notation $[[F]]$ indicates the difference between the values assumed by the quantity F on the two different sides of the boundary.
This condition implies that 
\be
[[\mathbf{n} \cdot \mathbf{B}]]=0,\;\; [[\mathbf{n} \cdot \mathbf{J}]]=0 .
\ee
where $\mathbf{J}$ is the current, and $\mathbf{n}$ is a unit vector perpendicular to the boundary.
The continuity of the tangential component of the electric field $\mathbf{E}$ reads
\be\label{boundary_ele}
0=[[\mathbf{n} \times \mathbf{E}]]=-[[\mathbf{n} \times \alpha \mathbf{B}]]+[[\mathbf{n} \times \eta \nabla\times \mathbf{B}]] .
\ee
In our case \eqref{boundary_ele} reads
\be
0=-\alpha_{-} (\mathbf{n} \times \mathbf{B})_{-}+\eta_{-}(\mathbf{n} \times \nabla\times \mathbf{B})_{-}
-\eta_{+} \beta_{+} (\mathbf{n} \times \mathbf{B})_{+}
\ee
which can be satisfied as long as  $\eta$ is not assumed 
to be infinite in the $r>R$ domain  (see P. H. Roberts in \cite{proctor}).

At last, after some algebra it is possible to show that the continuity of the field components  
determines $B_{nm}$, $C_{nm}$ and $D_{nm}$ as a function of $A_{nm}$ as follows:
\begin{subequations}
\label{zaal}
\ba
&&B_{nm}=\frac{A_{nm} (\calpha -\cbeta ) J_{n+\frac{1}{2}}(\calpha )}{\calpha  \cbeta }\\ 
&&C_{nm}=\frac{\pi A_{nm}}{2} \left(J_{n+\frac{1}{2}}(\calpha ) J_{n+\frac{3}{2}}(\cbeta )-J_{n+\frac{3}{2}}(\calpha)J_{n+\frac{1}{2}}(\cbeta )\right)\\
&&D_{nm}=\frac{\pi A_{nm}}{2} \left(J_{n+\frac{3}{2}}(\calpha ) Y_{n+\frac{1}{2}}(\cbeta )-J_{n+\frac{1}{2}}(\calpha) Y_{n+\frac{3}{2}}(\cbeta )\right)
\ea
\end{subequations}
We exclude the possibility that  $\calpha=\cbeta$, as in this case the interior field 
would also be force-free.
It is worth noticing that the surface toroidal field at the surface is in general non-zero, which was instead a necessary condition in the current-free case.

Whilst in the current-free case, as well as in the vertical (i.e. purely radial) field case, the ``quantization" condition for $\calpha$ is obtained by imposing the vanishing of  the toroidal field on the stellar surface \citep{krause80}, in this case the discrete turbulent spectrum is obtained by imposing the outer boundary condition at $r=\ro$.

In particular, following \cite{bo16}, it is assumed that at $r=\ro$ the 
radial component of the field is the dominant one, so that
\be
\label{obound}
B_\theta(r=\ro)=B_\phi(r=\ro)=0 . 
\ee
This is consistent with the presence of a stellar wind in the solution found by \cite{parker58}.

It is not difficult to show that \eqref{obound} implies 
the following equation to hold
\ba\label{quant}
&&0= Y_{n+\frac{1}{2}}(\ro \beta ) \left(J_{n+\frac{1}{2}}(\calpha ) J_{n+\frac{3}{2}}(\cbeta )-J_{n+\frac{3}{2}}(\calpha ) J_{n+\frac{1}{2}}(\cbeta)\right) 
+\nonumber\\
&&J_{n+\frac{1}{2}}(\ro \beta ) \left(J_{n+\frac{3}{2}}(\calpha )
   Y_{n+\frac{1}{2}}(\cbeta )-J_{n+\frac{1}{2}}(\calpha ) Y_{n+\frac{3}{2}}(\cbeta )\right) .
\ea
Clearly the eigenvalues for a given $n$ but different $m$ coincide, i.e. there is degeneration with respect to $m$ as
\eqref{quant} does not depend on $m$.
It is interesting to study the ``quantized" spectrum for various values of $\beta$  and $\ro$. In fact, while in the limit
$\beta\rightarrow 0$ it is possible to show that \eqref{quant} reproduces the well-known textbook solution for the
current-free case (see \cite{krause80}), in general the $\beta$-dependence modifies the standard current-free solution. 

\begin{figure}
\begin{center}
\includegraphics[width=0.5\textwidth]{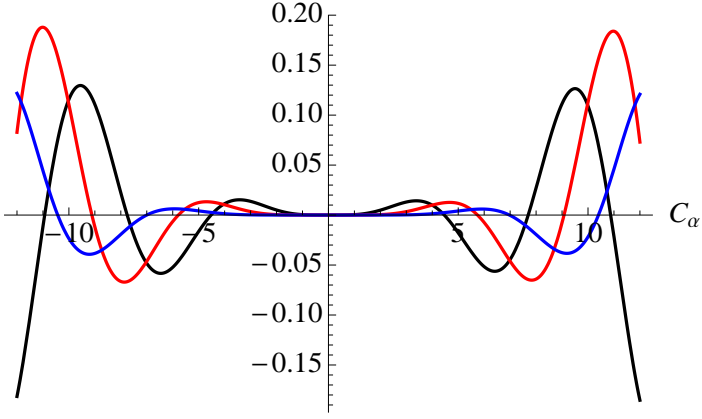}
\caption{Various zeroes of equation \eqref{quant} are displayed for $\ro=2 R$, $\cbeta=0.1$ 
and $n=1$ (black), $n=2$ (red) and $n=3$ (blue), respectively. The normalization is given by $A_{nm}=1$.
\label{fig1}}
\end{center}
\end{figure}

\begin{figure}
\begin{center}
\includegraphics[width=0.5\textwidth]{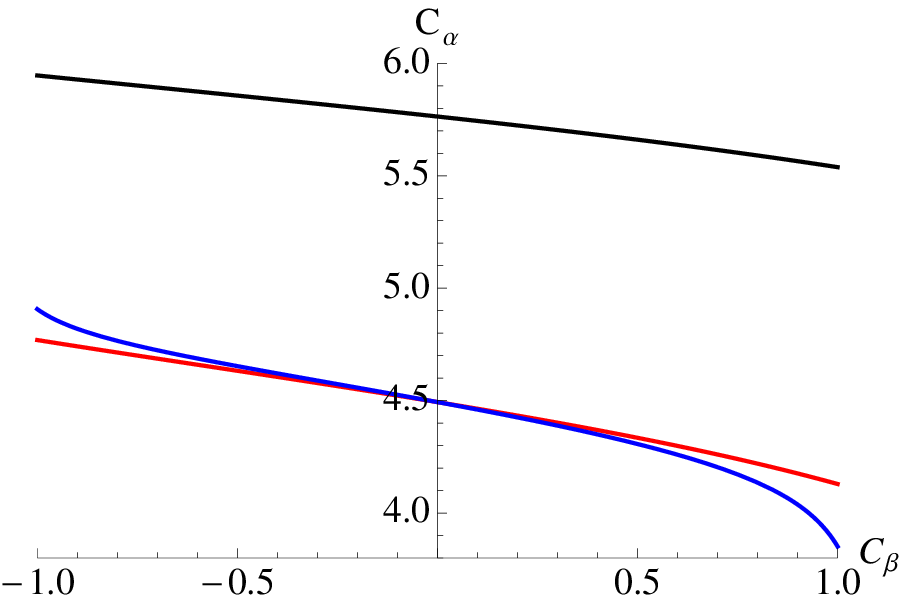}
\caption{Eigenvalues for the $n=2$ (black) and for the $n=1$ mode (red) as a function of $\cbeta$ for $\ro=2 R$. In blue the $n=1$ eigenvalue
is depicted for $\ro=3R$. \label{fig2}}
\end{center}
\end{figure}

\begin{figure}
\begin{center}
\includegraphics[width=0.5\textwidth]{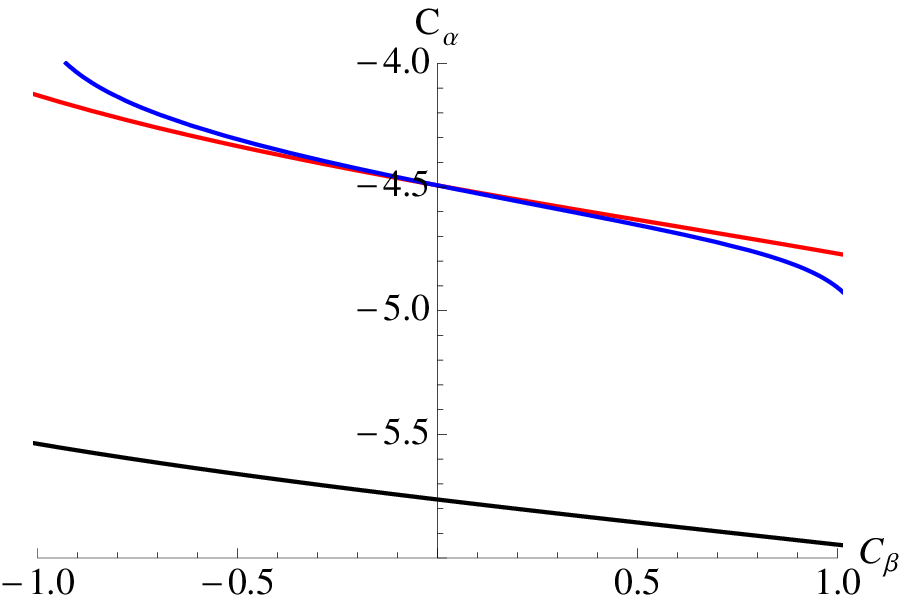}
\caption{Eigenvalues for the $n=2$ (black) and for the $n=1$ mode (red) as a function of $\cbeta$ for $\ro=2 R$ for negative values
of $\calpha$. In blue the $n=1$ eigenvalue
is depicted for $\ro=3R$. \label{fig3}}
\end{center}
\end{figure}

In \fig{fig1} the zeroes of \eqref{quant} are displayed for $\beta=0.1$ and $\ro=2 R$ and various values of $n$, while
in \fig{fig2} and \fig{fig3} the dependence of  $\calpha$ on $\cbeta$ is displayed for $n=1$ (red) and $n=2$ (black), whilst the blue
line represents the $n=1$ solutions with $\ro=3 R$.

It is  clear that $\calpha$ is a decreasing function of $\cbeta$ 
and this patterns is substantially unchanged for any value of $n$.
Moreover a positive $\cbeta$ produces values of $\calpha$ smaller than the current-free case, while negative values of 
$\cbeta$ lead to greater values of $\calpha$. In other words if the current in the corona is parallel to the field,
the dynamo is easily excited, while if the current is anti-parallel the dynamo condition is more difficult 
to attain.

Explicit values for $n=1$ are displayed in table (\ref{tab1}).

\begin{figure}
\begin{center}
\includegraphics[width=0.5\textwidth]{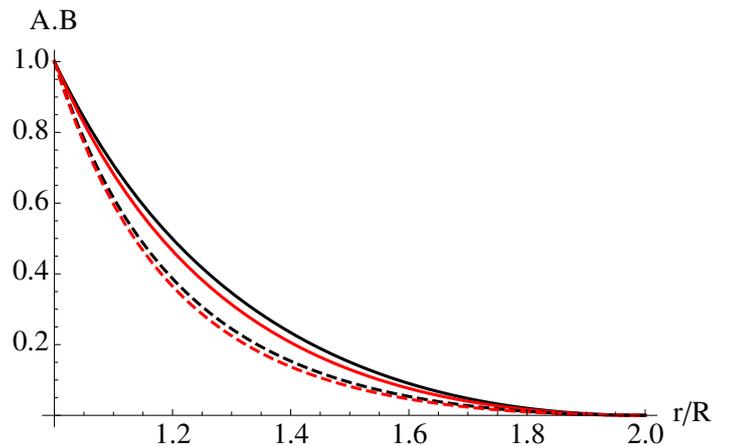}
\caption{Radial dependence of the magnetic helicity $\mathbf{A} \cdot \mathbf{B}$ 
 for the $n=1$ mode (solid line) mode for $ C_\beta =0.8$ (black) and $C_\beta =0.1$ (red)
and for the $n=2$ mode (dashed lines) at $\theta=\pi/4$. The magnetic helicity is normalized by using $A_{nm}=1$.
\label{fig4}}
\end{center}
\end{figure}

\begin{figure}
\begin{center}
\includegraphics[width=0.5\textwidth]{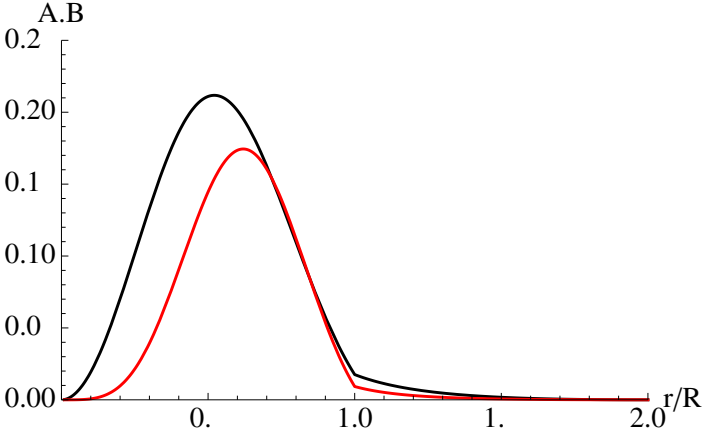}
\caption{Global radial dependence of the angle averaged magnetic helicity 
 $\int\mathbf{A} \cdot \mathbf{B}\, d\theta d\phi$ 
 for the $n=1$ mode (black) mode and for the $n=2$ mode (red) for $ C_\beta =0.5$ (arbitrary units).  
Note the presence of a small tail of non-zero helicity extending up to $r=1.5 R$.  The magnetic helicity is normalized by using $A_{nm}=1$.
 \label{fig5}}
\end{center}
\end{figure}

\begin{table}
\caption{The first four positive and negative ``eigenvalues" for $n=1$ for $\calpha$
for various values of $\cbeta$. Note the invariance of the spectrum with respect to a 
$\calpha\rightarrow -\calpha$ and $\cbeta\rightarrow -\cbeta$ change.} 
\centering                         
\begin{tabular}{l l l l l  }       
\hline\hline
\\[-8pt]         
$\cbeta$ & $C_\alpha^{11}$  & $C_\alpha^{12}$ & $C_\alpha^{13}$ & $C_\alpha^{14}$ \\    
\hline
\\[-2pt]
1 & 4.1290 & 7.3873 & 10.5752 & 13.7418 \\[2pt]
0.1 & 4.6384 & 7.6958 & 10.8748 & 14.0369 \\[2pt]
0.1 & -4.5222 & -7.7542 & -10.9331 & -14.0953 \\[2pt]
0 & 4.4934 & 7.7253 & 10.9041 & 14.0662 \\[2pt]
-0.1 & 4.5222 & 7.7542 & 10.9331 & 14.0953 \\[2pt]
-0.1 & -4.6384 & -7.6958 & -10.8748 & -14.0369 \\[2pt]
-1 & 4.7693 & 8.0139 & 11.1986 & 14.3639 \\[2pt]
\hline                                
\hline                                
\end{tabular}
\label{tab1}
\end{table}

The solution presented in this article allows in principle to explicitly compute 
the magnetic helicity flux across the stellar boundary, which is non zero for a non-vanishing 
$\cbeta$. Indeed the volume-integrated magnetic helicity 
$\mathbf{A}\cdot \mathbf{B}$ it is maximum at the inner boundary and it decays up to zero at the outer boundary as it is shown in \fig{fig4} and \fig{fig5}.   
This is not a coincidence: it is possible to to show that our outer boundary condition  \eqref{obound} 
amounts to the condition of no helicity flux across the outer boundary.

\subsection{Magnetic field components}
We can provide an explicit expression for the magnetic field in the outer part of the domain.

Reality condition requires $A_{n,m}={A}^\ast_{n,-m}$ in \eqref{zaal}. Therefore  \eqref{solff} 
reads
\ba
\label{solff2}
&&\Phi= R \sum_{n=1}^\infty\sum_{m=0}^{2 n}[a_{n}^m Q_n J_n{(\beta r)}\cos m\varphi\nonumber\\
&&+b_{n}^m  S_n Y_n(\beta r)\sin m \varphi] \left (\frac{R}{r} \right )^{1/2} P^m_n(\mu)
\ea
where, for $\beta>0$,  $a^m_n$ and $b^m_n$ are arbitrary real coefficients, and
\ba
&&Q_n=J_{n+\frac{1}{2}}(\calpha ) J_{n+\frac{3}{2}}(\cbeta)-J_{n+\frac{3}{2}}(\calpha)J_{n+\frac{1}{2}}(\cbeta)\\
&&S_n=J_{n+\frac{3}{2}}(\calpha ) Y_{n+\frac{1}{2}}(\cbeta)-J_{n+\frac{1}{2}}(\calpha)Y_{n+\frac{2}{2}}(\cbeta) .
\ea
Moreover $P^m_n(\mu)$ are the  normalized associated Legendre functions of degree $n$, order $m$  and argument $\mu=\cos \theta$. 
In particular, $\calpha$ and $\cbeta$ are not arbitrary but are linked via the quantization 
condition given by  \eqref{quant}.

We thus have
\ba
&&B_r(r,\theta,\varphi)= \nonumber\\
&&\left ( \frac{R}{r} \right )^{3/2}
\sum_{n=1}^\infty\sum_{m=0}^{2 n} \left[ a_n^m Q_n n (n+1) J_{n+\frac{1}{2}}(\beta r)\cos m\varphi \right. \nonumber \\
&& \left. +b_n^m S_n Y_{n+\frac{1}{2}}  (\beta r) \left (n(n+1)-\frac{m^2}{\sin^2\theta} \right ) \sin m\varphi \right ] P^m_n(\mu) 
\ea
\ba
&&B_\theta (r,\theta,\varphi)=\nonumber\\ 
&&\frac{1}{2}\left ( \frac{R}{r} \right )^{3/2}
\sum_{n=1}^\infty\sum_{m=0}^{2 n} \left[ - 2 a_n^m Q_n  r \beta  J_{n+\frac{1}{2}}(\beta r) \frac{m}{\sin \theta} \sin m \varphi P^m_n(\mu)  \right. \nonumber \\
&&  +a_n^m Q_n [J_{n+\frac{1}{2}}(\beta r)+2 r \frac{d}{dr} J_{n+\frac{1}{2}}(\beta r)] \cos m \varphi \frac{d}{d \theta} P^m_n(\mu)\nonumber\\
&& \left. +b_n^m S_n [Y_{n+\frac{1}{2}}(\beta r)+2 r \frac{d}{dr} Y_{n+\frac{1}{2}}(\beta r)] \sin m \varphi \frac{d}{d \theta} P^m_n(\mu)\right ]
\ea
\ba
&&B_\phi (r,\theta,\varphi)=\nonumber\\ 
&&-\frac{1}{2}\left ( \frac{R}{r} \right )^{3/2}
\sum_{n=1}^\infty\sum_{m=0}^{2 n} \left[  2 a_n^m Q_n [J_{n+\frac{1}{2}}(\beta r) \right.\nonumber\\
&&+2 r \frac{d}{dr} J_{n+\frac{1}{2}}(\beta r)]  \frac{m}{\sin \theta}\sin m \varphi P^m_n(\mu)
+\left (2 \beta r a_n^m Q_n J_{n+\frac{1}{2}}(\beta r) \cos m \varphi \right. \nonumber\\
&&\left. \left. +b^m_n S_n Y_{n+\frac{1}{2}}(\beta r) \sin m \varphi \right ) \frac{d}{d\theta} P^m_n(\mu) \right ]
\ea

\section{Conclusion}
The analytical linear force-free solution presented in this paper has been obtained by coupling 
a corona with a dynamo generated field in the interior. 
Although it is a very idealized situation, it shows several interesting features.
The most important property of the solution is the endowment
of a new dependence of the dynamo number on the strength and 
the topology of the force-free field as parametrized by the parameter $\beta$.  Positive $\beta$ produce smaller dynamo
numbers, while negative $\beta$ render the dynamo more difficult to excite. This is in agreement with the 
harmonic atmosphere model in \cite{bo16}, as Beltrami fields are also harmonic, 
whilst the converse is not true in general.
The toroidal field is non-zero at the surface and therefore it could be important to implement this solution in ZDI regularization procedure. 
Indeed as the coupling with the interior has significantly reduced the number of free parameters needed to specify the field  for each
harmonics, the space of possible solutions could be significantly reduced.  We hope to discuss possible physical application
of our solution in a forthcoming paper, where we will extend our approach to the non-stationary dynamo case.


\emph{Acknowledgments}

We thank Axel Brandenburg for comments and useful discussions.
FDS acknowledges the Swedish Research Council International Postdoc fellowship for support, Giuseppe Jacopo Guidi for discussions and INAF Astrophysical observatory of Catania for warm hospitality that allowed this work to be carried out.

\bibliographystyle{apj}
\bibliography{hh}
\label{lastpage}


\end{document}